
\magnification = 1200
\hsize = 6.3  true in
\vsize = 8.3 true in
\baselineskip = 15pt plus 2pt   
\parskip = 6pt			

\font\tentworm=cmr10 scaled \magstep2
\font\tentwobf=cmbx10 scaled \magstep2
\font\tenonerm=cmr10 scaled \magstep1
\font\tenonebf=cmbx10 scaled \magstep1
\font\eightrm=cmr8
\font\eightit=cmti8
\font\eightbf=cmbx8
\font\eightsl=cmsl8
\font\sevensy=cmsy7
\font\sevenm=cmmi7

\font\twelverm=cmr12
\font\twelvebf=cmbx12
\def\subsection #1\par{\noindent {\bf #1} \noindent \rm}

\def\mid {\let\rm=\tenonerm \let\bf=\tenonebf \rm \bf}

\def\para{\par \vskip 12 pt}

\def\head{\let\rm=\tentworm \let\bf=\tentwobf \rm \bf}

\def\heading #1 #2\par{\centerline {\head #1} \smallskip
 \centerline {\head #2} \vskip .15 pt \rm}

\def\eight{\let\rm=\eightrm \let\it=\eightit \let\bf=\eightbf
\let\sl=\eightsl \let\sy=\sevensy \let\m=\sevenm \rm}

\def\foots{\noindent \eight \baselineskip=10 true pt \noindent \rm}
\def\sexion{\let\rm=\twelverm \let\bf=\twelvebf \rm \bf}

\def\section #1 #2\par{\vskip 20 pt \noindent {\mid #1} \enspace {\mid #2}
  \para \noindent \rm}

\def\ssection #1 #2\par{\noindent {\mid #1} \enspace {\mid #2}
  \para \noindent \rm}

\def\abstract#1\par{\para \foots {\bf Abstract: \enspace}#1 \para}

\def\author#1\par{\centerline {#1} \vskip 0.1 true in \rm}

\def\abstract#1\par{\noindent {\bf Abstract: }#1 \vskip 0.5 true in \rm}

\def\midsection #1\par{\noindent {\sexion #1} \noindent \rm}

\def\sqr#1#2{{\vcenter{\vbox{\hrule height#2pt
 \hbox {\vrule width#2pt height#1pt \kern#1pt
  \vrule width#2pt}
  \hrule height#2pt}}}}


\def \m {|\mu|}
\def \map
\def \nm {\nu + \m}

\def \half {1\over 2}

\def \apj {Astrophys. J}
\def \apjl {Astrophys. J. Lett.}

\def \prl {Phys. Rev. Lett.}

\def \mnras {Mon. Not. R. Astr. Soc.}
\def \aa {Astron. Astrophys.}
\def \nat {Nature}
\def \doublespace {\baselineskip = 20pt plus 7pt \message {double space}}
\def \singlespace {\baselineskip = 13pt plus 3pt \message {single space}}
\singlespace

\def\heading #1 #2\par{\centerline {\head #1} \smallskip
 \centerline {\head #2} \vskip .15 pt \rm}

\def \body {\vfill \eject \parindent = 1.0 true cm
	\ifx \spacing \singlespace \singlespace \else \doublespace \fi}

\def \title#1 {\centerline {{\bf #1}}}
\def \Abstract#1 {\noindent \baselineskip=15pt plus 3pt \parshape=1 40pt310pt
  {\bf Abstract} \ \ #1}








\catcode`@=11
\def \C@ncel#1#2 {\ooalign {$\hfil#1 \mkern2mu/ \hfil $\crcr$#1#2$}}
\def \gf#1 {\mathrel {\mathpalette \c@ncel#1}}	
\def \Gf#1 {\mathrel {\mathpalette \C@ncel#1}}	

\def \gapx {\;\lower 2pt \hbox {$\buildrel > \over {\scriptstyle {\sim}}$}
\; }
\def \lapx {\;\lower 2pt \hbox {$\buildrel < \over {\scriptstyle {\sim}}$}
\; }


\topskip 3.3 true cm

\footline = {\ifnum \pageno = 1 \hfill \else \hfill \number \pageno \hfill \fi}
\def\half{{1\over 2}}
\def\dr{{\delta \rho \over \rho}}
\def\dk{\delta_k}
\def\Dk{\delta _k ^2}
\def\ms{M_{\odot}}
\def\dt{{\Delta T\over T}}
\def\ie{\it{ ie}}
\def\n{\noindent}

\def\v2{\vskip 0.2 true cm}
\def\v3{\vskip 0.3 true cm}


\line{\hfill \foots{\bf IUCAA Preprint 02-92}}
\line{\hfill \foots{ Jan. 1992}}
\vskip .45 true in

\centerline{\bf {The Large Scale Structure of the Universe: Theory {\it vs.}
Observations} \footnote*{Invited talk delivered at the International Conference
on General Relativity and Cosmology (ICGC), Ahmedabad, India;
13 - 18 December 1991. To appear
in its proceedings. }}

\vskip 1.0cm \centerline{\bf{Varun Sahni}}
\vskip 0.5cm \centerline{Inter-University Centre for Astronomy and
Astrophysics}
\centerline{Post Bag 4, Ganeshkhind, Pune 411 007, India}
\bigskip
\topskip 1.0cm


\n \centerline {I.\phantom{.}\bf Introduction}
\vskip .4cm
Observations indicate that the Universe is homogeneous and isotropic
on scales $\ge$ 100 - 200 Mpc., whereas on smaller scales, its fundamental
units
-- galaxies -- cluster together to form groups, clusters and even
superclusters,
 with effective mass scales ranging from $10^{12}\ms$ (groups) to $10^{15}\ms$
(superclusters).
There are also indications that most of the Universe consists of large empty
regions -- voids, which are virtually devoid of any matter.
Although many statistical indicators are used to describe the observed
clustering of
galaxies$^{29}$, the best researched and perhaps most robust indicator of
galaxy
clustering is the two point galaxy - galaxy correlation function $\xi (r)$, and
its angular counterpart $w(\theta)$.
Recent investigations$^1$ indicate that $w(\theta)$
remains positive till  $\sim 50 - 100 h^{-1}$ Mpc. indicating that galaxies
continue to cluster even on such large scales (see fig. 1).
(We have chosen the Hubble
parameter to be $H_0 = 100\times h km.s^{-1} Mpc^{-1})$.
The decreasing amplitude of $w(\theta)$ indicates that clustering is getting
weaker with scale, so that one is justified in assuming the Universe to be
fairly homogeneous on scales $\ge 100h^{-1}$ Mpc.

\n
A detailed knowledge of the observed clustering of galaxies, as well as their
gravity induced random motions, provides us with a sensitive test against
which to probe models of structure formation. In this review, I shall attempt
to confront theory with observations in order to try and constrain models of
galaxy
formation, dwelling in some detail on the currently popular Cold Dark Matter
model.

\n
I shall mainly concern myself with the following
three sets
of observations with which to try and constrain theory:

1) The observed isotropy of the Cosmic Microwave Background Radiation,

2) The clustering of galaxies measured by the two point correlation function
$\xi (r)$,

3) The large scale peculiar velocity field of galaxies.

\n
\centerline {II.\phantom{.}\bf The Cosmic Microwave Background Radiation}

A remarkable feature of the Cosmic Microwave Background Radiation (CMBR) is
that it is accurately described by a thermal distribution of photons with
temperature$^2$ $\simeq 2.7 ^\circ K$ (Fig 2a).
The CMBR also appears to be
smooth to better than one part in ten thousand on virtually all angular scales
(after one has subtracted out a visible dipole component caused presumably
by the motion of our galaxy with respect to the Hubble flow).

The observational upperlimits
on the CMBR anisotropy are summarised in fig. 2b, the most stringent
constraints on $\dt (= {{T - T_0}\over T_0}, T_0 \approx 2.7^\circ K)$
arising on arc minute scales are :
$\dt < 1.5\times 10^{-5}$ at $\theta = 7.15'$, and are provided by observations
 at the Owens Valley Radio Observatory$^3$ (OVRO).
Since a comoving scale $\lambda$ subtends an angle $\theta = 34.4'' (\Omega h)
\times \lambda_{Mpc} \equiv 65.4'' (\Omega^{{2\over 3}} h^{{1\over 3}})
({M\over
10^{12} M_{\odot}})^{1\over 3}$, it follows that masses of the order
$10^{13}\ms \le
M(\theta) \le 10^{17}\ms$ -- covering the entire mass range from rich groups to
superclusters of galaxies$^4$ are covered by angular scales ranging between
: few arc min
$\le \theta \le 3^\circ$.
As a result, the absence of fluctuations in the
CMBR on these angular scales constrains
the form of the density perturbation spectrum on scales
of $2 - 100$ Mpc, and is consequently an important check on theories of
galaxy formation.\footnote*{
Since recombination is not instantaneous, fluctuations on very small angular
scales $\le \Delta\theta \sim 10 (\Omega h^2)^{-\half}$
arc min, are wiped out because of the finite
thickness of the last scattering surface .}

On scales $\ge 3^\circ$ the main contribution to the CMBR arises because of
primordial gravitational perturbations on the surface of last scattering
${\delta T \over T} \simeq {1\over 3} {\delta \phi \over c^2}$
( the {\it Sachs-Wolfe} effect). Anisotropies on such large scales when
detected
are likely to be particularly revealing because they probe scales that were
larger
 than the Horizon size at recombination. Consequently, any information
concerning
 the
microwave anisotropy on such scales, also carries with it implicit
information regarding the primordial form of the fluctuation spectrum
which might have been
generated by physical processes occurring at the very beginning of the big
bang, such as inflation.

\bigskip
\vskip .4 cm
\centerline {III.\phantom{.}\bf Non-baryonic models of galaxy formation}

\vskip .4cm
Since most scenario's of galaxy formation predict a definite, albeit small
amplitude of density fluctuations at the surface of last scattering, present
observational constraints on $\dt$ effectively rule out a host of cosmological
models including many in which structure formation proceeds via gravitational
instability, such as baryonic models of galaxy formation, the hot dark matter
model with an early epoch of galaxy formation, and a low density cold dark
matter model$^5$. Models which are ruled out also include those in which
structure forms nongravitationally due to the propogation of shocks driven
by primordial explosions$^6$ or by the collapse of superconducting cosmic
strings$^7$.
In the latter class of models, a sizeable amount of energy is injected into
the intergalactic medium during the formation of shocks, leading to large
distortions in the microwave spectrum which have not been observed.

Models which survive the stringent CMBR constraints include the standard
Cold Dark Matter (CDM) model,
as well as the cosmic string scenario of galaxy
formation. Since Alex Vilenkin will discuss string models of galaxy formation
at this meeting, I shall mainly focus on reviewing the present status of the
CDM model {\it vis a vis} observations.

The Cold Dark Matter model presupposes that most of the matter in the Universe
today, exists in the form of nonrelativistic ($\it {ie}.$ cold), weakly
interacting, non - baryonic matter, which decoupled from the rest of
the baryonic matter
in the Universe at some early epoch, soon after the big bang.

The need for incorporating nonbaryonic forms of matter into the standard
big bang model, arises because of the necessity to generate large enough
perturbations by the present epoch, to account for the presence of galaxies,
from sufficiently small initial values of the density contrast $\dr$,
without violating the CMBR constraints on small -- arc minute -- scales.
This is clearly a difficult task in baryonic models in which perturbation
growth takes place only after the cosmological recombination of hydrogen,
which occurs at redshifts $\sim 1100$.
(Before recombination, radiation pressure caused by Thompson scattering,
effectively prevents the growth of all perturbations having wavelenths
smaller than $\sim 180 h^{-1} $Mpc. -- the horizon size at recombination).
After recombination, density perturbations grow linearly with the scale factor
of the Universe -- $a(t)$, until the Universe
becomes curvature dominated,  after which time perturbation growth is strongly
suppressed $^{20}$ (if $\Omega \le 1$).

The maximum permitted growth for perturbations is therefore
${\delta\over\delta_{rec}} \leq {a_0\over a_{rec}} \simeq 1100$
[for $\Omega \le 1, \Omega \equiv {\rho\over \rho_{cr}} =
{8\pi G \rho\over 3 H^2}$ where $\rho$ is the present mass density of the
Universe.]
Consequently, the requirement that $\dr > 1$ today -- which is necessary for
the
formation of gravitationally bound systems such as galaxies -- leads to the
primordial amplitude: $\dr \approx 10^{-3}$ at recombination.
If the primordial perturbations are adiabatic
($\ie$ fluctuations for which the specific
entropy per baryon is spatially constant), then at recombination$^{4,8}$
$$
(\dr)_B = {3\over 4}(\dr)_{\gamma} = 3\dt
\eqno(1)
$$
so that a fluctuation amplitude $\sim 10^{-3}$ in the baryon component would
invariably result in an anisotropy $\sim 10^{-4}$ in the CMBR temparature on
arc minute scales which has not been observed. On the other hand, perturbations
 in a nonbaryonic component, can begin growing soon after matter - radiation
equality has been achieved ($\ie$ by $z \sim 10^4$ in the case of an $\Omega
= 1$ Universe$^9$,) with the result that large final perturbations can develop
from small initial values without violating the CMBR constraints on small
scales. Fig. 3 shows how perturbations in baryons and in a nonbaryonic
component grow after matter-radiation equality; it is clear that
whereas perturbations in the non-baryonic component can grow soon after
matter dominance, the growth in the baryonic component is suppressed until
after recombination. (After recombination, baryons catch up with the
perturbations
in the non-baryonic component, by falling into the potential wells
created by the non-baryonic perturbations.)
Consequently, since the CMBR anisotropy in (1) is
determined by the density fluctuation in the baryonic component
at recombination, it is an order
of magnitude smaller than it would have been in a purely baryonic scenario,
and therefore does not come into conflict with observations.

Historically, the suggestion that a significant portion of the matter in the
Universe may exist in the form of weakly interacting massive particles
(WIMP's) such as massive neutrino's, was made by Marx and Szalay$^{10}$ as
well as by Cowsik and McClelland$^{11}$. Galaxy formation with massive
neutrino's
began to be studied in earnest after Lubimov claimed to have detected a mass
of $\sim$ 30 eV for the electron neutrino$^9$. (The results of this work have
however, since been called into question).
Perturbations in a medium made up of light collisionless particles such as
massive
neutrino's, are subject to collisionless phase mixing , ({\it ie.} weakly
interacting particles free stream relativistically from regions of high
density into low density regions), which effectively wipes out perturbations
on scales smaller than $\lambda_{fs} \sim 40$ Mpc$ {30eV\over m_\nu}$ --
corresponding to scales of clusters of galaxies$^{12}$. \footnote*{
$\lambda_{fs} $ is the distance traversed by a relativistic massive neutrino
until its momentum becomes non-relativistic}. Consequently the perturbation
spectrum in a massive neutrino model (conveniently called the Hot Dark Matter
model -- to highlight the relativistic nature of its particle species),
displays a sharp cutoff at the free - streaming scale $\lambda_{fs} $, which
is shown in figure 4.   The presence of a cutoff on scales $\sim 40 Mpc$
ensures that the first objects to undergo gravitational collapse in this
model will be of cluster scales , with smaller-scale objects forming
out of the subsequent fragmentation of cluster-sized pancakes. Such a
scenario for galaxy formation was originally suggested by Zeldovich in the
70's in connection with baryonic models of galaxy formation, and is popularly
known as the {\it pancake} or {\it top-down} model of galaxy formation$^{13}$.
Some characteristic features of an HDM Universe which emerge from N-body
simulations$^{14,15}$, include the presence of a cell-like structure with
sharply
demarkated filaments and voids on scales of roughly $\sim \lambda_{fs}$ Mpc .

Although the gross features of the HDM model do seem to reproduce several
aspects of the observed large scale structure in the Universe -- such as the
superclusters and voids seen in redshift surveys, the model when tested
quantitatively
fails to agree with observations.
N-body computer simulations of Hot Dark Matter models$^{14,15}$ for instance,
show that in order to agree with such statistical indicators of galaxy
clustering as the observed two point galaxy-galaxy correlation function at the
 present
epoch , galaxy formation in this model would have to have taken place very
recently ($z \le 1$), which clearly runs counter to observations of
galaxies and quasars at redshifts $z \ge 3$. Although some attempts are
still on for
a revival of this model$^{16}$, most theorists have since the mid 80's,
begun to favour a different non-baryonic model for galaxy formation -- the Cold
Dark Matter (CDM) model.

The non-baryonic particle candidates in CDM models are assumed to be  weakly
interacting and nonrelativistic (hence "cold"). Particle candidates which fit
this prescription are known to arise naturally in particle physics models
incorporating supersymmetry, such as the supersymmetric partners of the
photon and
graviton -- the photino and gravitino, respectively.
Others, such as the axion, are required to provide consistency to low energy
theories such as QCD$^{17}$.

Due to the absence of free streaming, the fluctuation spectrum for CDM --
$\Dk$ --
\footnote{**}{($\dr) = {1\over (2\pi)^3}\int{\dk\exp(k^{\mu} x_{\mu})d^3k}$}
does not show a cutoff on small scales$^{18,19}$, instead $\Dk$ falls off
monotonically, approaching the asymptotic form $\Dk \propto k^{-3}\log ^2 k$
for
$k >> k_{eq}$ (see (2) below) ($ k = {2 \pi a(t) \over \lambda}$ is the
comoving
 wavenumber).

The CDM spectrum is well described by the the semi-analytic
approximation obtained by Starobinsky and Sahni$^{19, 21}$
$$
\Dk = {1\over A^4k^3} {ln^2(1 + Bk)\over \bigl( 1 + {ln(1 + Bk)\over (Ak)^2}
\bigr)^2}
\eqno(2)
$$
where $A = 3.08 \sqrt\kappa h^{-2} Mpc;
B = 1.83 \sqrt\kappa h^{-2} Mpc; \kappa = {\Omega_{rel}\over \Omega_\gamma}.$
(See also ref.[44]).

The characteristic bend in the CDM spectrum occurs at length scales $\sim
\lambda_{eq}$ ($= {13\over \Omega h^2}$ Mpc), corresponding to the Horizon
scale
at matter - radiation equality. (This is the only length scale that enters
into the spectrum). This bend arises because density perturbations on scales
much smaller than $\lambda_{eq}$ enter the horizon when the Universe was still
radiation dominated. Since perturbations in radiation get ironed
out on scales smaller than the Hubble radius (due to the free streaming of
photons),
 matter
perturbations have nothing to gravitate towards, and so grow at the exceedingly
slow rate
: $\delta \propto 1 + {3\over 2} {a(t)\over a_{eq}}$ until matter
dominance
$(a = a_{eq})$. On the other hand, larger than horizon size perturbations
continue to grow as $a^2(t)$,
while the Universe is radiation dominated , with the result that
modes entering the horizon at later times come in with correspondingly
larger amplitudes, resulting in a bend in the shape of $\Dk $.
Perturbations on scales $> \lambda_{eq}$, reenter the horizon after matter
dominance and so are left effectively untouched by
the radiation era. As a result, the fluctuation spectrum preserves its
primordial form $\Dk \propto k$, on scales $ > \lambda_{eq}$ .

A measure of the density contrast over a given scale is provided by
$\dr \sim (k^3|\Dk |)^{\half}$, which grows as $\ln^2 k$ on small
scales$^{19,21}$ (see fig. 4).
Consequently, smaller scales are the first to go  non-linear , with larger
scales following suit.
As a result gravitational clustering takes place
hierarchically with large units such as clusters and superclusters ,
forming out of the merger of smaller -- galaxy and globular cluster-sized
units. This
hierarchical process of building larger structures out of smaller ones, is
often
referred to as the {\it bottom-up} scenario of galaxy formation, and is
complementary to the
{\it top-down} scenario which is associated with the HDM model.
The {\it rms} fluctuation in
the mass found within a randomly placed sphere of radius $R: {\Delta M\over M}$
is plotted in fig. 5 for the CDM spectrum given in (2). From fig. 5 and the
relation
${{ \Delta M \over M}(0,R)\over { \Delta M \over M}(z,R)} \simeq 1 + z$,
it is easy to show that objects of mass $\sim 10^{11} - 10^{12}\ms$ can begin
forming by redshifts $\sim 3 - 4$, in the standard (unbiased), Cold Dark Matter
model ($1+z = {a_0\over a}$).

Due to the fact that non-baryonic matter is dissipationless and does not shine,
 CDM models can
successfully account for the existence of spherically symetric dark halo's
surrounding galaxies, and also for the substantial quantity on non-luminous
matter in clusters whose existence can be  inferred from arguments based on
the virial theorem. Other successes of the CDM model -- demonstrated using
N-body
 simulations -- include its ability to
account for the masses, sizes, angular momenta and abundance of
galaxies$^{22}$.

The CMBR anisotropies expected in a CDM Universe are$^5$:
$\dt (7.15') \simeq {10^{-5}\over b}$ and $\dt (7^\circ)
\simeq {10^{-5}\over b}$,
which are smaller than the observational upperlimits on these scales
($b$ is the biasing factor; $h = 0.5$ is assumed for the Hubble parameter).
For an unbiased CDM model (b = 1), the predictions for $\dt$ on large scales
are close to being either confirmed or ruled out by COBE .\footnote*{
Just as this paper was nearing completion, a press release
announced the detection of an anisotropy in the microwave background
measured by COBE$^{46}$:
$\dt \simeq 1.1\times 10^{-5}$, on scales $> 7^\circ$, which is consistent
with the predictions made by the standard (unbiased) CDM model.}

The varied diversity of its successes made the CDM model very popular in the
aftermath of the demise of the HDM scenario. However, recent observations of
galaxy clustering on large scales$^{1,23}$ ($50 - 100$ Mpc) -- fig. 1, as
well as measurements of the
peculiar velocities of galaxies$^{24}$ on scales $\sim 50 h^{-1}$ Mpc, have
posed a
serious observational challenge to the CDM scenario and have caused
considerable
debate in the cosmology community as to the viability of this model.
Some of these observational tests will be discussed in the following two
sections of this paper.
\vskip .4cm
\centerline {IV.\phantom{.}\bf Clustering of Galaxies}
\vskip .4cm

Galaxy clustering is most evident in wide angle survey's of the sky both in
the optical as well as in the infrared and 21cm wave bands (see fig 6).
Optical redshift surveys have reconfirmed this result, discovering in
addition, large-scale coherent structures such as the Perseus-Pisces
supercluster chain, and the great void in Bootes$^{25,26}$ ( fig 7).

The clustering of galaxies can be characterised by several statistical
indicators, the best known and most comprehensively studied being
the two point galaxy - galaxy correlation function -- $\xi (r)$, defined as
the probability in excess of random, of finding a galaxy at a distance {\it r}
from another, randomly picked galaxy .

On small scales ($r < 10 h^{-1}$ Mpc) the correlation function has the well
known form
$$
\xi(r) = ({r\over r_0})^{-1.8}
\eqno(3a)
$$
where $r_0 = 5 h^{-1}$ Mpc, is the clustering scale.

Since $\xi(r)$ is the fourier transform of the power spectrum $\Dk$
$$
\xi(r) = {1\over (2 \pi)^3}\int{|\Dk|\exp(k^{\mu}{x_\mu})d^3k},
\eqno(3b)
$$
it carries direct information of the form and amplitude of the density
fluctuation spectrum. On large scales ($> \lambda_{eq} \approx
{13\over \Omega h^2}$ Mpc), $\Dk$ is not yet distorted by non-linear
gravitational
clustering, so that for primordial spectra of the form $\dk \propto k^n,$
  $\xi(r) \approx - \sin({\pi n\over 2})\times r^{- (3+n)};$
therefore, knowing the sign of
$\xi(r)$ and its slope on linear scales, one can directly infer the
primordial form of $\dk$.
Recent estimates  of the angular galaxy - galaxy correlation function --
$w(\theta)$, \footnote*{$w(\theta)$ and $\xi(r)$ can
be related via the Limber equation$^4$, so that for power law
spectra $\xi(r) \propto
r^{-\gamma}, w(\theta) \propto \theta^{1-\gamma}$.}
made using the Automatic Plate Measuring Machine (APM) survey, and covering
over 2 million galaxies, indicate that galaxies
continue to be positively clustered upto scales $\sim (50 - 100) h^{-1}$ Mpc
(see fig 1). Comparison of $w(\theta)$
with predictions of the standard CDM
model, shows that galaxies appear to be more strongly clustered on large
scales than can be accounted for in a $\Omega = 1$ CDM model with a
conventional Harrison - Zeldovich spectrum: $\Dk \propto k$ on large scales.
(The observational curve for $w(\theta)$ is consistent with a primordial
spectral index $-1 \le n \le 0$, but inconsistent with n = 1.)

Other indications of the existence of greater power on large scales comes from
recent redshift surveys of infrared galaxies taken with the help
of the IRAS satellite$^{23}$. These surveys show an $\it {rms}$. fluctuation of
$\sim 0.3$
in the number of galaxies on scales $\sim 20h^{-1}$ Mpc, which is about 2 - 3
times the predicted amplitude on these scales in a biased CDM scenario.
\footnote{**}{Light does not trace mass in biased galaxy formation scenario's,
consequently density fluctuations in these models, are smaller by a factor
$b = {{\delta N\over N}\over {\delta M\over M}}$ , than in an
unbiased scenario. (${\delta N\over N}$ is the variance in the number
density of bright galaxies on a given scale,
${\delta M\over M}$ is the variance in the mass, evaluated on the same scale.)}

There are several ways of circumventing the above problems with CDM.
One is to drop the assumption that the primordial perturbation spectrum is
of the scale invariant Harrison - Zeldovich type. Although a scale invariant
spectrum arises naturally in most models of inflation, models do
exist$^{31,32}$
in which the inflationary expansion of the Universe is of the power - law kind:
$a(t) \propto t^p, p > 1$, which leads to a more general spectrum for density
fluctuations$^{33,34}$: $\Dk \propto k^n$ where $ n = {3-p\over 1-p}$.
Such models
predict greater power on large scales for $\Dk$, and consequently also for
$\xi(r)$ and $w(\theta)$.

Another possibility for reconciling the observed evidence for large scale
clustering, is by working with low density CDM models ($\Omega \approx 0.1$
being preferred from estimates of the mass to light ratio in clusters of
galaxies). In such models the turnaround point in the CDM power spectrum (2),
occurs at $\lambda_{eq} \sim {13\over \Omega h^2}$ Mpc $\sim 130$ Mpc, so that
on scales $\sim 50 Mpc$, the slope of the perturbation spectrum
$\Dk\propto k^n$ is effectively $n \approx -1$ which is consistent with
observations of $w(\theta)$ on these scales.
However low density CDM Universes predict large distortions in the CMBR
and are ruled out by
observations$^5$. One can get around this difficulty by introducing an
effective
cosmological constant into the model,\footnote*{Although a small value of the
cosmological constant does not conflict with observations, its introduction
necessitates some fine tuning which most cosmologists find unattractive.}
$\Lambda = 3H_0^2 (1-\Omega_m)$ , so that
$\Omega_\Lambda + \Omega_m = 1$.
Such a cosmological model is consistent with the observed isotropy of the
microwave background,
predicts the correct slope and amplitude for $w(\theta)$ on large scales,
and in addition, is old
enough to explain the existence of the oldest observed star clusters in our
galaxy$^{35}$ (whose ages lie in the range $15 \le T \le 18$ billion years, and
are
difficult to accomodate within the framework of a matter dominated $\Omega = 1$
cosmogony$^{36}$.)
\vskip .4cm
\centerline {V.\phantom{.}\bf Large-Scale Peculiar Velocities of Galaxies}
\vskip .4cm

The cosmic microwave background radiation defines for all practical purposes,
an absolute frame of reference against which departures from a smooth Hubble
flow may be measured\footnote{**}{If {\bf V} is the observed velocity of a
galaxy
located at a distance {\bf r}, then its peculiar velocity is {\bf v} =
{\bf V } - H {\bf r}.}. An indication that our galaxy is not at rest with
respect to this reference frame, comes from the presence of a dipole
anisotropy in the CMBR having the form:
$$
T(\theta) = T_0 (1 + {v\over c} \cos\theta)
\eqno(4)
$$
where $v$ is the velocity of the observer, and $\theta$ is the angle at which
the microwave temperature is being measured with respect to the motion of the
observer. The {\it dipole anisotropy} -- indicating that the sky appears hotter
in
one direction and colder in the opposite one -- has the well established
value $ {\delta T\over T} = 1.2\times 10^{-3}$, and implies (after correcting
for the motion of the Earth) a velocity for the sun of $V^{sun}_{CM}
= 360 \pm 25$ km/s. Since the Suns motion in the Milky Way ($\sim 250$ km/s),
is in a direction roughly opposite to that of its motion with respect to
the CMBR,
we get $V^{gal}_{MB} = 540 \pm 50 $ km/s, for the motion of our galaxy
relative to the microwave background. Taking into account the relative motions
of galaxies within the local group, (which constitutes $\sim 20$ members, the
largest being the Milky Way and M31) we finally obtain $V^{LG}_{MB} =
610 \pm 50$ km/s, for the velocity of the barycentre of the local group
with respect to the CMBR.

The nearest large concentration of mass in the vicinity of the local group
is the Virgo cluster, located at a distance of $\sim 13 h^{-1}$ Mpc. The
relative peculiar velocity of the local group with respect to the Virgo
cluster (commonly known as {\it infall towards Virgo}) is
$V^{LG}_{Virgo} \approx
250$ km/s., indicating that the Virgo cluster contributes only
{\it partially} to
our overall peculiar motion. Furthermore, the direction of motion of the Local
Group relative to the CMBR, is roughly at an angle of $45^\circ$ {\it away}
from
the direction of the Virgo cluster, indicating that the Virgo cluster --
with the Local Group at its periphery \footnote*{constituting a patch
$\sim 13h^{-1}$ Mpc. in scale}
-- is moving at an overall velocity of $\sim 400$ km/s., in
the direction of the Hydra - Centaurus supercluster.
For sometime it was felt that perhaps the gravitational attraction towards
Hydra -
Centaurus might wholly account for the large peculiar velocity of Virgo and
the Local Group (see fig 8). However work on $\sim 400$ elliptical galaxies
by Lynden-Bell and others$^{24}$, revealed that large peculiar motions were
not confined to the Local Group alone, but were shared by galaxies occupying
large volumes of space $\sim 50 h^{-1}$ Mpc., and probably extending to
regions well beyond the Hydra - Centaurus supercluster.

Simple models have attributed most of the bulk motions $(600 \pm 100)$ km/s,
to the presence of a large mass concentration $(\sim 10^{16}\ms)$ located at
a distance $\sim 42 h^{-1}$ Mpc. away from us and behind the Hydra - Centaurus
supercluster, which has appropriately been dubbed -- The Great Attractor.
As of today, attempts to find the great attractor are on, and it is hoped
that conclusive evidence of its existence -- such as a reversal of galaxy
infall on the far side of the Great Attractor$^{45}$ -- will soon be
forethcoming.

The existence of large scale peculiar motions of $\sim 600 \pm 100$ km/s on
scales $\sim 50 h^{-1}$ Mpc. has provided a stiff challenge to theories of
galaxy formation, especially the Cold Dark Matter model. One can estimate the
predicted value of bulk motions in a hierarchical theory of structure formation
such as CDM, by noting that since most of the {\it action} is on large scales,
it is safe to use linear theory to provide an estimate of the peculiar
velocity of galaxies on such scales. Following this procedure, we note that
a peculiar acceleration $\bf a_p $
acting for a time $\tau$, will induce a peculiar velocity ${\bf v} = {\bf a_p}
\times \tau$, where ${\bf a_p}$ is related to the  over all mass
enhancement on
scales $r : {\bf a_p} = {G\delta M {\bf r}\over r^3} = {4\over 3} \pi G \delta
\rho {\bf r} = {\Omega H\over 2} {\bf v_H} (\dr)$, where ${\bf v_H} = H{\bf r}$
is the Hubble velocity.
Since$^4$ ${3\over 2}H\Omega \tau \simeq \Omega^{0.6}$ we obtain:
$$
\bigl({{\bf v}\over {\bf v_H}}\bigr)_{\lambda} =
{\Omega^{0.6}\over 3} (\dr)_{\lambda}
\eqno(5)
$$
for the relative peculiar velocity on a scale $\lambda$.

\n Eq. (5) enables us to relate the peculiar velocity on a given scale with the
corresponding value of the density enhancement on that scale. Thus on the
scale of the Great Attractor
$(\lambda \sim 50 h^{-1}$ Mpc), ${{\bf v}\over {\bf v_H}} \sim
0.1$ which gives $(\dr)_{50} \simeq 0.3$, for an $\Omega = 1 $ Universe.
If we assume that the density perturbation spectrum on a given scale can be
characterised by a power law: $\Dk \propto k^n$, then the density contrast
on that scale is given by $(\dr)_{\lambda} \sim (k^3|\Dk|)^{\half} \sim
\lambda^{-{3+n\over 2}} $. In order to yield sensible information,
the above expression has first to be normalised to agree with the observed
variance in the number counts of bright galaxies on a given scale.
Requiring$^8$
$(\dr)_8 = {\delta N\over N}(8h^{-1} Mpc) = 1$ gives
$$
(\dr)_{\lambda} = ({\lambda\over 8 h^{-1}})^{-{3+n\over 2}}
\eqno(6)
$$
so that on scales of the Great Attractor,
$(\dr)_{50} = ({8\over 50})^{{3+n\over 2}}$.

In an $\Omega = 1$ CDM model, the spectrum $\dk$ aquires its primordial
Harrison - Zeldovich form $\Dk \propto k$ on scales $\ge 50$ Mpc, so that
substituting $n = 1$ in (6) we find $(\dr)_{50} \simeq 0.03$, which is an
order of magnitude smaller than the value inferred from bulk flows on these
scales. Since $\dr$ is a Gaussian on large scales, the probability of finding
a density enhancement $\sim \delta_{GA}$ on scales of $\sim 50 h^{-1}$ Mpc
is given by
$$
P(\delta \ge \delta_{GA}) = {1\over \sqrt{2\pi} \delta_{exp}}\int_{\delta_{GA}}
^{\infty}{exp{-({\delta\over 2 \delta_{exp}})^2} d\delta}
\eqno(7)
$$
where $\delta_{exp}$ is the expected value of $\dr$ on a given scale.
Substituting $\delta_{exp} \simeq 0.03$ and $\delta_{GA} \simeq 0.3$ in (7)
we get $P(\delta \ge \delta_{GA}) \simeq 5\times 10^{-24}$ -- {\ie} the
probability of finding a region like the great attractor in a standard
CDM Universe is extremely small$^{41,42}$.

In our discussion of galaxy clustering in the previous section, we mentioned
that the CDM model could be reconciled with the observed large amplitude
of the angular correlation function $w(\theta)$ if the spectral
index on scales $50 - 100$ Mpc. was less than unity -- the preferred value
being $n \simeq -1$.
(This possibility arises for flat CDM models dominated by a
cosmological constant. )
Substituting $n = -1$ in (6) we find $(\dr)_{50}
\simeq 0.16$ and the corresponding probability of finding a region like the
great
attractor is now $P(\delta \ge \delta_{GA}) \simeq 0.04$, which although
still small, is not at all unlikely.

One can redo the above calculation for determining the velocity perturbation
in an expanding Universe more accurately, by using the linearised
continuity equation:
$$
{\dot \delta} = {1\over a} \nabla {\vec v}
\eqno(8)
$$
Fourier transforming (8) we get: ${\dot \delta_k} =
{1\over a}i{\vec k}{\vec v_k};$ since ${{\dot \delta_k}\over \delta_k}
= {d\log(\delta(t))\over d\log(a(t))} H(t) \simeq \Omega^{0.6}H(t)$
we finally obtain
${\vec v_k} = -iH\Omega^{0.6}{\delta_k{\vec k}\over k^2}$.
The $\it rms$ value of the bulk velocity on a scale $R$ can now be calculated
by choosing a suitable window (filtering) function $W(kR)$, to filter out the
small scale contribution, so that finally
$$
\langle v^2 \rangle = {1\over (2\pi)^3}\int{d^3k |v_k|^2 W(kR)}
= {(H \Omega^{0.6})^2\over 2\pi^2}\int {|\Dk| exp(-{(kR)^2\over 2}) dk}
\eqno(9)
$$
(where we have used a Gaussian filter: $W(kR) = exp(-{(kR)^2\over 2})$).

The probability of measuring a peculiar velocity having an $\it rms$ value --
$v_{rms}$ in the interval $v_1 \le v \le v_2$ is given by$^8$
$$
P(v) = \sqrt{{54\over \pi}} \int_{v1}^{v2}{({v\over v_{rms}})^2 exp{-{3\over 2}
({v\over v_{rms}})^2} {dv\over v_{rms}}}
\eqno(11)
$$
{}From (11) it follows that the probability of measuring a velocity $v$ in the
range ${v_{rms}\over 3} < v < 1.6\times v_{rms}$ is $\sim 90\%$.
Applying (9) and (11) to the CDM model with $\Dk$ given by (2) we obtain:
$v_{rms}(r \simeq 50 h^{-1} Mpc) \simeq 83 km s^{-1} h^{-0.92}$, so that, at
the
90\% confidence level $30 km/s < v(r=50h^{-1}) h^{-0.92} < 135 km/s $, which
once again demonstrates that high peculiar velocities on scales $\sim 50
h^{-1}$Mpc
 are extremely difficult to accomodate within the framework of an
$\Omega = 1$ CDM model.

Assuming $\Dk \propto k^n$ in (9) we obtain : $v(r) \propto r^{-{n+1\over 2}}$.
For an $\Omega = 1$ CDM model $\Dk \propto k$ on scales $ \sim 50 h^{-1}
$Mpc, giving
$v(r) \propto r^{-1}$ on large scales. On the other hand for low $\Omega$
CDM models, $n \simeq -1$ on scales $ \sim 50 h^{-1}
$Mpc, which implies
$v \approx constant$.

A comparison of $v(r) \propto r^{-{n+1\over 2}}$ with $\dr(r) \propto
r^{-{3+n\over 2}}$, shows that large scales are weighed much more
heavily in $v(r)$ than in $\dr(r)$.
A quantity having still greater power on large scales than either $v(r)$ or
$\dr$, is the peculiar gravitational potential, which is related to $\dr$
via the Poisson equation: $\Delta \phi = 4\pi G a^2 \delta \rho$.
For a spherically symmetric distribution of mass --
$$
\phi(r) \simeq {G \delta M \over r} \simeq {4\over 3} \pi G \rho (a(t) r)^2 \dr
\eqno(10)
$$
where $a(t) r = R$, is the physical
length
scale, and $\rho$ -- the background density of the Universe.
Since $\rho \propto a^{-3}(t)$, and $\dr \propto a(t)$ ,
we find that fluctuations in the linear gravitational potential are independent
 of time in an
Einstein - de Sitter Universe. From (10) we also find that $\phi(r) \propto
r^{-{n-1 \over 2}}$, which demonstrates that $\phi(r)$ is
{\it scale invariant} for a primordial Harrison - Zeldovich spectrum (n = 1).

\vskip .4cm
\centerline {VI.\phantom{.}\bf Loitering Cosmological Models}
\vskip .4cm
In the preceeding two sections we have been mainly dealing with models in
which non-barionic dark matter plays the key role in determining how and where
structure forms in the Universe. The reason for excluding baryonic models
from our consideration was two-fold. Firstly (as discussed in $\S 2$),
baryonic models suffer from the {\it growth problem}: too little growth in
density perturbations occurs in baryonic models, to account for the presence
of galaxies by the present epoch. Secondly, in order to be consistent with
inflation, one requires that $\Omega = 1$, whereas primordial nucleosynthesis
strongly suggests$^{17}$ $\Omega_b \le 0.04 h^{-2}$.

One would also like to point out, that in addition to these problems faced
primarily by baryonic models of structure formation, all matter dominated
cosmological models with $\Omega = 1$, suffer from the so-called
{\it age problem}, {\it ie.} the Universe turns out to be too young
to accomodate a population of "old" objects, such as globular
clusters$^{36}$.

In the remainder of this paper I will discuss a scenario for galaxy formation,
in which only baryons cluster, and in which both the growth, as well as the
age problems are absent.
According to this scenario -- dubbed {\it loitering Universe} ,
the Universe underwent a recent phase of very slow expansion (or loitering),
during which
its scale factor was either a constant, or grew {\it very slowly} with
time$^{39}$.
The advantages of such an expansion regime are two-fold: firstly, a Universe
which loitered in the past can be much older than a matter dominated
Einstein - de Sitter Universe whose age $T \simeq {2\over 3 H} \le 13$ billion
years for values of the Hubble parameter
$ 100 \ge H \ge 50 km s^{-1} Mpc^{-1}$
indicated by observations. (This is illustrated in fig(9).)

Secondly, since inhomogeneities grow quasi-exponentially during
{\it loitering} (see fig(10)), galaxies can form by the present epoch out very
small initial fluctuations. Therefore a low density of matter which clusters
is not a hinderence to the growth of perturbations. Furthermore, any feature
in the density power spectrum associated with the transition from radiation-
domination to matter-domination (such as $\lambda_{eq}$ in $\S 3$)
is on a larger scale than in an Einstein-
deSitter Universe,
which might be the explanation for the extra power in $w(\theta)$
which is observed .

These features of a loitering Universe have been discussed in detail
by Sahni, Feldman and Stebbins$^{39}$, who also show that if one limits
the density in baryons to $\Omega_b \simeq 0.2$ (in keeping with virial
estimates of dark matter in clusters of galaxies), then loitering could only
have occured in a fairly narrow redshift interval $2 \le z \le 7.2$.
Thus it may be possible to observationally rule out (or confirm) loitering,
by  studying the number count of galaxies as a function of their redshift,
as well as by observations relating to the number density of absorption lines
in
the spectra of distant QSO's. Some work in this direction has been reported
in this conference by Patrick Das Gupta.
\vskip .4cm
\centerline {\bf Discussion}
\vskip .4cm
In this brief review I have attempted to confront realistic models of structure
formation with observational data describing the properties of our Universe on
scales $\ge 20 h^{-1}$ Mpc. On such scales density fluctuations are still
linear and it is possible to make a systematic check of theoretical predictions
of cosmological models {\it vs.} observations, without resorting to
detailed nonlinear calculations such as N-body simulations.

We have seen that although the CDM model is very successful in explaining the
small scale texture of galaxy clustering, it seems to lack sufficient large
scale
power to explain either the clustering of galaxies on scales $\sim 50 - 100$
Mpc., or to successfully account for the large observed bulk motions
of galaxies on scales $\sim 50 h^{-1}$ Mpc.
As alternatives to the {\it standard} matter dominated CDM model, both a CDM
model with a cosmological constant, as well as the loitering Universe model
of $\S 6$, seem to have greater success in predicting
more power on larger scales, as required by observations.

{\bf Note.} Just as this paper was nearing completion, a press release
announced the detection of an anisotropy in the microwave background
measured by COBE$^{46}$:
$\dt \simeq 1.1\times 10^{-5}$, on scales $> 7^\circ$ (the
associated quadrupole being $Q \simeq 5\times 10^{-6}$). The COBE results
indicate a spectrum $\Dk \propto k^n$, with $n = 1.1 \pm 0.6$ on scales
$> 700 h^{-1} $ Mpc. ($n = 1$ is the scale-invariant Harrison - Zeldovich
spectrum.)
The anisotropies in the CMBR measured by COBE, are consistent with the
unbiased, matter dominated, CDM model, as well as with a CDM model with a
non-zero cosmological constant. Most other models of structure formation
however (with the exception of scenarios involving cosmic strings or
a loitering epoch) seem to be ruled out.

\vfill\eject

\centerline {\bf Acknowledgements}
\vskip .4cm
I am grateful to J.V. Narlikar for bringing the COBE preprints$^{46}$
to my notice. I also thank Arvind Paranjpye for help in preparing the
manuscript.
\vskip 1.4cm
\centerline {\bf  References}
\vskip 1.4cm
\n
[1] Maddox, S.J., Efstathiou, G., Sutherland, W.J. and Loveday, J.,
\mnras {\bf 242} (1990) 43.
\vskip .2 cm
\n
[2] Mather, J. et al., \apj {\bf 354} (1990) L37.
\vskip .2 cm
\n
[3] Readhead, A.C.S. et al., \apj {\bf 346} (1989) 566.
\vskip .2 cm
\n
[4] Peebles, P.J.E., {\it The Large Scale Structure of the Universe}
(Princeton University, Princeton (1980);
Zeldovich, Ya.B., and Novikov, I.D., {\it The Structure and Evolution of the
Universe} (University of Chicago, Chicago/London 1983)
\vskip .2 cm
\n
[5] Bond, J.R., in: {\it The Early Universe} eds: W.G. Unruh and G.W.
Semenoff (D. Reidel Publishing Company 1988); Bond, J.R., CITA Preprint (1990).
\vskip .2 cm
\n
[6] Ostriker, J.P., and Cowie, L., \apj {\bf 243} (1981) L127.
\vskip .2 cm
\n
[7] Ostriker, J.P., Thompson, C. and Witten, E., Phys. Lett. B {\bf 180}
(1986) 231.
\vskip .2 cm
\n
[8] Silk, J. and Vittorio, N., in: {\it Confrontation between Theories and
Observations in Cosmology: Present Status and Future Programs} --
Proceedings of the International School of Physics {\it Enrico Fermi}.
eds. J. Audouze and F. Melchiorri (North Holland 1990)
\vskip .2 cm
\n
[9] Shandarin, S.F., Doroshkevich, A.G. and Zeldovich, Ya.B.,
Sov. Phys. Usp. {\bf 26} (1983) 46.
\vskip .2 cm
\n
[10] Marx, G. and Szalay, A.S., in Proceedings Neutrino (1977) 123
(Technoinform, Budapest 1977)
\vskip .2 cm
\n
[11] Cowsik, R. and McClelland, J., \apj {\bf 180} (1973) 7.
\vskip .2 cm
\n
[12] Doroshkevich, A.G., Khlopov, M.Yu., Sunyaev, R.A., Szalay, A.S. and
Zeldovich, Ya.B., in: Proceedingsof the Tenth Texas Symposium on Relativistic
astrophysics, eds: R. Ramaty and F.C. Jones, Annals of the New York Academy
of Sciences {\bf 375} (New York academy of Sciences, New York 1981)

Bond, J.R. and Szalay, A.S., \apj {\bf 274} (1983) 443.
\vskip .2 cm
\n
[13] Zeldovich, Ya.B., \aa {\bf 5} (1970) 84.

Zeldovich, Ya.B., \mnras {\bf 160} (1972) 1p.
\vskip .2 cm
\n
[14] White, S.D.M., Frenk, C. and Davis, M., \apj {\bf 274} (1983) L1.

White, S.D.M., Frenk, C. and Davis, M., \apj {\bf 287} (1983) 1.
\vskip .2 cm
\n
[15] Centrella, J. and Melott, A., \nat {\bf 305} (1982) 196.
\vskip .2 cm
\n
[16] Braun, E., Dekel, A. and Shapiro, P.R., \apj {\bf 328} (1988) 34.
\vskip .2 cm
\n
[17] Kolb, E.W. and Turner, M.S., {\it The Early Universe}
(Addison - Wesley 1990)
\vskip .2 cm
\n
[18] Bond, J.R. and Efstathiou, G., \apjl {\bf 285} (1984) L45.
\vskip .2 cm
\n
[19] Starobinsky, A.A. and Sahni, V., in: {\it Modern Theoretical and
Experimental Problems of General Relativity } (MGPI Press, Moscow 1984) 77.

Sahni, V. PhD Thesis, Moscow State University (1984) (Unpublished).
\vskip .2 cm
\n
[20] Guyot, M. and Zeldovich, Ya.B., \aa {\bf 9} (1970) 227.

Meszaros, P., \aa {\bf 37} (1974) 225.

Sahni, V., Pramana {\bf 15} (1980) 423.
\vskip .2 cm
\n
[21] Shandarin, S.F. and Zeldovich, Ya.B., Rev. Mod. Phys. {\bf 61} (1989) 185.
\vskip .2 cm
\n
[22] Davis, M., Efstathiou, G.,Frenk, C. and White, S.D.M., Nature {\bf 356}
(1992) 489.
\vskip .2 cm
\n
[23] Saunders, W. et al., Nature {\bf 349} (1991) 32.
\vskip .2 cm
\n
[24] Lynden-Bell, D. et al., \apj {\bf 326} (1988) 19.
\vskip .2 cm
\n
[25] Giovanelli, R. and Haynes, M.P., \apj {\bf 292} (1986) 404.
\vskip .2 cm
\n
[26] Kirshner, R.P., Oemler jr., A., Schechter, P.L. and Schectman, S.A.,
\apjl {\bf 248} (1981) L57.

Kirshner, R.P., Oemler jr., A., Schechter, P.L. and Schectman, S.A.,
\apj {\bf 314} (1986) 493.
\vskip .2 cm
\n
[27] Koo, D., Kron, R. and Szalay, A., in: Proceedings of the XIII Texas
Symposium on Relativistic astrophysics, ed. M. Ulmer (World Scientific,
Singapore 1986) 284.
\vskip .2 cm
\n
[28] Joeveer, M., Einasto, J. and Tago, E., \mnras {\bf 185} (1978) 357.
\vskip .2 cm
\n
[29] Melott, A., Phys. Rep. {\bf 193} (1990) 1.
\vskip .2 cm
\n
[30] Geller, M.J., in: {\it Confrontation between Theories and
Observations in Cosmology: Present Status and Future Programs} --
Proceedings of the International School of Physics {\it Enrico Fermi}.
eds. J. Audouze and F. Melchiorri (North Holland 1990)
\vskip .2 cm
\n
[31] La, D. and Steinhardt, P.J., \prl {\bf 62} (1989) 376.
\vskip .2 cm
\n
[32] Berkin, A.L., Maeda, K. and Yokoyama, J., \prl {\bf 65} (1990) 141.
\vskip .2 cm
\n
[33] Lucchin, F., Matarrese, S. and Vittorio, N., \aa {\bf 162} (1986) 13.
\vskip .2 cm
\n
[34] Liddle, A.R., Lyth, D.H. and Sutherland, W., Phys. Lett. B {\bf 279}
(1992) 244.
\vskip .2 cm
\n
[35] Efstathiou, G., Sutherland, W.J. and Maddox, S.J., Nature {\bf 348}
(1990) 705.

Kofman, L.A. and Starobinsky, A.A., Sov. Astron. Lett. {\bf 11} (1985) 5.

Gorski, K.M., Silk, J. and Vittorio, N., \prl {\bf 68} (1992) 733.
\vskip .2 cm
\n
[36] Janes, K. and Demarque, P., \apj {\bf 264} (1983) 206.
\vskip .2 cm
\n
[37] Padmanabhan, T. and Subramanian, K., Bull. Astron. Soc. of India (In
press).
\vskip .2 cm
\n
[38] Kashlinsky, A., in: {\it Large-Scale Structures and Peculiar Motions in
the Universe} eds. D.W. Lathew and L.A. Nicolai da Costa (A.S.P. Conference
Series, vol. 15, 1991)
\vskip .2 cm
\n
[39] Sahni, V., Feldman, H. and Stebbins, A. \apj {\bf 385} (1992) 1.
\vskip .2 cm
\n
[40] Lahav, O., \mnras {\bf 225} (1987) 213.
\vskip .2 cm
\n
[41] Bertschinger, E. and Juskeweicz, R., \apjl {\bf 334} (1988) 59.
\vskip .2 cm
\n
[42] Kashlinsky, A. and Jones, B.J.T., Nature {\bf 349} (1991) 753.
\vskip .2 cm
\n
[43] Villela, T., Meinhold, P. and Lubin, P. in: {\it Confrontation
between Theories and
Observations in Cosmology: Present Status and Future Programs} --
Proceedings of the International School of Physics {\it Enrico Fermi}.
eds. J. Audouze and F. Melchiorri (North Holland 1990)
\vskip .2 cm
\n
[44] Bond, J.R. and Eftathiou, G., \apjl {\bf 285} (1984) L45.
\vskip .2 cm
\n
[45] Dressler, A. and Faber, S.M., \apjl {\bf 354} (1990) L45.
\vskip .2 cm
\n
[46] Smoot, G.F., et al. Submitted to \apjl (1992)

Wright, E.L., et al. Submitted to \apjl (1992)
\vskip .2 cm
\n
\vskip .4cm
\centerline {\bf Figure Captions}
\vskip .4cm
Fig. 1 The angular correlation function $w(\theta)$ for galaxies in the
Automatic Plate Measuring Machine (APM) galaxy survey$^1$ --
dots represent galaxies.
The CDM prediction is shown by the smooth curve.
\vskip .2 cm
\n
Fig. 2a Spectrum of the CMBR (taken with the FIRAS instrument on COBE$^2$)
compared to a blackbody with a temperature of $2.735 \pm 0.06 ^{\circ} K$.
Boxes are centered on measured points and show
a size of assumed error $\sim 1\%$. The units fot the vertical axis are
$10^{-4}ergs s^{-1} cm^{-2} sr^{-1}$ cm.
\vskip .2 cm
\n
Fig 2b. Anisotropy limits$^{2,3,43}$ on $\dt$. The solid line at $\theta >
7^{\circ}$ corresponds to the recent detection of a CMBR anisotropy:
${\Delta T\over T} \simeq 1.1\times 10^{-5}$ on scales $\theta > 7^{\circ}$
by COBE$^{46}$.
\vskip .2 cm
\n
Fig. 3 Growth of linear density perturbations$^9$.
The wavelength of the perturbation is taken to be $\simeq \lambda_{eq}$ --
the comoving horizon scale at matter - radiation equality ($\lambda_{eq}
\simeq {13\over \Omega h^2} Mpc$). The solid lines correspond to
perturbations in baryons -- $(\dr)_B$, and the dashed line to perturbations
in a non - baryonic component -- $(\dr)_X$, such as Hot or Cold Dark Matter.
The dotted vertical lines $z_{eq}$ and $z_{rec}$ correspond to epochs of
matter - radiation equality, and recombination, respectively.
\vskip .2 cm
\n
Fig. 4 The density fluctuation spectrum $\sqrt{k^3\times\Dk}$ is plotted for:
The Hot Dark Matter Model$^{12}$
and the Cold Dark Matter Model$^{19,21,44}$ .
An arbitrary normalisation is assumed.
\vskip .2 cm
\n
Fig. 5 The ${\it rms}$ mass fluctuation on a scale $R$, ${\Delta M\over M}(R)$
, is plotted
for an unbiased CDM Universe with $\Omega = 1$ and h = 0.5.
${\Delta M\over M}$ is normalised to agree with the observed excess in the
number density of galaxies at $8 h^{-1} Mpc$.
(${\Delta M\over M}^2(R,t_0) = {1\over 2\pi^2} \int{k^2 dk |\Dk |
W^2(kR)}$,
$W(kR) = 3 ({sin(kR)\over (kR)^3}
- {cos(kR)\over (kR)^2})$, is the {\it top hat} window function$^{17}$.)
\vskip .2 cm
\n
Fig. 6 Equal area projections of the a) North, and b) South, galactic
hemispheres,
taken with the IRAS satellite$^{40}$. The galactic plane lies along the
equator. The numbers along the circumference indicate longitude.
The major clusters in the Northern hemisphere include -- Virgo (centre) and
Hydra - Centaurus ($270^\circ - 310^\circ$). In the Southern hemisphere:
Persius - Pisces ($130^\circ - 150^\circ$),
Pavo - Indus ($315^\circ - 345^\circ$) and Fornax.
\vskip .2 cm
\n
Fig. 7 A CfA redshift slice$^{30}$ containing $\sim 1067$ galaxies. The
survey strip is centered at $\delta = 29.5^\circ$ and is
$6^\circ$ in declination. The observed velocity {\it vs.} right ascension
is shown.
\vskip .2 cm
\n
Fig. 8 Velocities for the Local Group and the Virgo cluster are shown in a
rough pen sketch (not drawn to scale !).

\n LG denotes the Local group, HC -- the Hydra - Centaurus supercluster, and
GA -- the region of the Great Attractor.
\vskip .2 cm
\n
Fig. 9 The scale factor in the loitering Universe scenario$^{39}$. The early
evolution is matter dominated $a \propto t^{2\over 3}$ followed by a loitering
phase $a \simeq constant$. As can be seen, the Universe is {\it very} old,
{\it ie}\phantom{.}  $T_0 >> H_0^{-1}.$
\vskip .2 cm
\n
Fig. 10 Growth of linear inhomogeneities ${\delta(t)\over \delta_i}$
superimposed
over the scale factor $a(t)$ evolution for a loitering Universe. In this figure
${\delta\over \delta_i}$ grows by a factor of $10^5$. (Scales for the two
curves
are different.)
\vskip .2 cm
\n

\vfil\eject

\end